# Boron monosulfide: equation of state and pressure-induced phase transition

K.A. Cherednichenko,[1] I.A. Kruglov,[2,3] A.R. Oganov,[3,4] Y. Le Godec,[5] M. Mezouar,[6] and V.L. Solozhenko [1,*]

[1] *LSPM–CNRS, Université Paris Nord, 93430 Villetaneuse, France*

[2] *Dukhov Research Institute of Automatics (VNIIA), Moscow, 127055, Russia*

[3] *Moscow Institute of Physics and Technology, Dolgoprudny, Moscow Region, 141700, Russia*

[4] *Skolkovo Institute of Science and Technology, Skolkovo Moscow Region, 143026, Russia*

[5] *IMPMC, UPMC Sorbonne Universités, CNRS – UMR 7590, 75005 Paris, France*

[6] *European Synchrotron Radiation Facility, 38043 Grenoble, France*

## Abstract

*Quasi-hydrostatic compression of rhombohedral boron monosulfide (r-BS) has been studied up to 50 GPa at room temperature using diamond-anvil cells and angle-dispersive synchrotron X-ray diffraction. A fit of the experimental P–V data to the Vinet equation of state yields bulk modulus $B_0$ of 42.2(1.4) GPa and its first pressure derivative $B_0'$ of 7.6(2) that are in excellent agreement with our ab initio calculations. Formation of a new high-pressure phase of boron monosulfide (hp-BS) has been observed above 35 GPa. According to ab initio evolutionary crystal structure predictions combined with Rietveld refinement of high-pressure X-ray diffraction data, the structure of hp-BS has trigonal symmetry and belongs to the space group P-3m1. As it follows from electron density of states calculations, the phase transformation is accompanied by an insulator-metal transition.*

## I. Introduction

The $A^{III}B^{VI}$ semiconductors (GaS, GaSe, InS, InSe, etc.) are promising materials for solar cells, nonlinear optics, photovoltaic energy converters, radiation detectors, photoresistors and solid-state batteries [1-3]. Rhombohedral boron monosulfide, *r*-BS, is a light-element member of this family. Unlike other $A^{III}B^{VI}$ semiconductors, *r*-BS still remains poorly understood, especially under pressure [4].

Very recently, the phonon properties of *r*-BS have been studied experimentally and theoretically up to 34 GPa at room temperature [5]. Based on the results of corresponding *ab initio* LCAO calculations the equation-of-state (EOS) parameters of *r*-BS ware estimated i.e. $B_0$ = 31.2 GPa and $B_0' = 8.4$. However, the theoretical EOS data have not been supported experimentally so far. In the present work, equation of state and phase stability of rhombohedral BS have been studied up to 50 GPa by synchrotron X-ray powder diffraction.

---

[*] e-mail: vladimir.solozhenko@univ-paris13.fr

## II. Experimental

Polycrystalline rhombohedral BS was synthesized at 7.5 GPa and 2200 K by reaction of amorphous boron (Johnson Matthey, 99%) and sulfur (Johnson Matthey, spectrographic grade) according to the method described elsewhere [5]. The structure and phase purity of the samples were confirmed by powder X-ray diffraction (G3000 TEXT Inel diffractometer, Cu$K\alpha$1 radiation) and Raman spectroscopy (Horiba Jobin Yvon HR800 spectrometer, 632.8-nm He-Ne laser). The lattice parameters ($a = 3.0586$ Å, $c = 20.3708$ Å) of the synthesized $r$-BS are in good agreement with literature data [4].

High-pressure experiments in the 2-50 GPa range were carried out at room temperature in Le Toullec type membrane diamond anvil cells [6] with 300 µm culet anvils. $r$-BS powder was loaded into 100-µm hole drilled in a rhenium gasket pre-indented down to ~30 µm. Neon pressure transmitting medium was used in all experiments. The sample pressure was determined from the shift of the ruby R1 fluorescence line [7] and equation of state of solid neon [8]. Pressure was measured before and after each X-ray diffraction measurement and further the mean pressure values were used, the pressure drift during experiment did not exceed 0.5 GPa; the maximum pressure uncertainty was 1 GPa.

*In-situ* X-ray diffraction studies have been performed at ID27 beamline, European Synchrotron Radiation Facility and Xpress beamline, Elettra Sincrotrone Trieste. Patterns were collected in angle-dispersive mode with a monochromatic beam; the wavelengths were selected equal to 0.3738 Å at ID27 beamline and 0.4957 Å at Xpress beamline. In both cases the MAR 345 image plate detectors were employed for data acquisition. The exposure time varied from 20 seconds at ID27 beamline to 600 seconds at Xpress beamline. The collected diffraction patterns were processed using Fit2D software [9]. The unit cell parameters at various pressures have been refined using PowderCell [10] and Maud [11] software; the values are presented in Table I.

## III. Computational methodology

Evolutionary metadynamics method [12,13] was used as the algorithm in search for high-pressure phase(s) of BS. Through the use of history-dependent potential this method allows a system to cross energy barriers and sample new low-energy structures. It leads to the exploration of new reaction pathways and acceleration of observation of new crystal structures. Metadynamics method provides a reach list of low-enthalpy metastable phases and gives insight into phase transition mechanism. Since a new BS phase was synthesized at high pressure and room temperature, the use of evolutionary metadynamics allows us to find both thermodynamically stable and kinetically accessible structures. We used the USPEX code [12-14] which combines the advantages of classical metadynamics and evolutionary structure prediction, and proved itself in many scientific cases [15-17]. Rhombohedral $R$-$3m$ structure of BS was used as a starting point, the calculation was carried out at 42 GPa for 50 generations, each generation consisted of 25 structures.

Energy calculations were performed using density functional theory (DFT) as implemented in the Vienna *Ab initio* Simulation Package (VASP) [18] within the generalized gradient approximation of Perdew-Burke-Ernzerhof [19]. Projector augmented wave (PAW) [20] method was employed to describe core electrons. We used plane wave energy cutoff of 500 eV and Γ-centered k-point mesh with the resolution of 2π·0.05 Å$^{-1}$. The optimized exchange van der Waals functional (optB86-vdW) [21,22] was also applied to take into account van der Waals interactions in the system. The new phase of boron monosulfide (see Table III) appeared in the 3$^{rd}$ generation. We also performed variable-composition calculation using evolutionary algorithm USPEX in the B-S system at 50 GPa. A detailed description of the crystal structure of the new phase is presented in section IV.B.

## IV. Results and Discussion

### A. Equation of state of r-BS

Rhombohedral boron monosulfide is known to have trigonal symmetry with the space group *R*-3*m*. Its unit cell in the hexagonal setting contains three layers with A-B-C stacking motif (Fig. 1*a*). One layer of *r*-BS consists of the chain of trigonal $S_3B$-$BS_3$ anti-prisms, so that the B-B pairs aligned along the *c*-axis are sandwiched between hexagonal layers of S atoms, rotated by π/3 relative to each other. There are two different bond types in rhombohedral BS: strong ionic-covalent intralayer bonds and weak van der Waals interlayer bonding along *c* axis. Due to this fact, the anisotropic compression of *r*-BS unit cell is expected.

Fig. 2 presents experimental data on *r*-BS lattice compression along different unit-cell axes. The data obtained at ID27 and Xpress beamlines are in excellent agreement. Due to this, here and further we will considered the integral dataset. As expected, the compression along the *c*-axis is more significant than along the *a*-axis due to considerable shrinking of the weak interlayer bonds. To approximate the nonlinear relation between the lattice parameters and pressure the one-dimensional analog of Murnaghan equation of state (Eq. 1) has been used following the work [23]:

$$r = r_0 \left[1 + P \left(\frac{\beta'_0}{\beta_0}\right)\right]^{-\frac{1}{\beta'_0}} \quad (1)$$

Here *r* is the lattice parameter (index 0 refers to ambient pressure); $\beta_0$ is 300-K axial modulus and $\beta'_0$ is its first pressure derivative. The $\beta_{0,a}$ and $\beta_{0,c}$ axis moduli that fit best all experimental data are 431.3±8.4 GPa and 47.1±1.5 GPa, respectively. The pressure derivatives are $\beta'_{0,a}$ = 10.7(7) and $\beta'_{0,c}$ = 14.2(3). The axial moduli can be easily transferred in the linear compressibilities ($k_r$) according to Eq. 2:

$$k_r = \beta_{0,r}^{-1} = \left(\frac{d \ln(r)}{d P}\right)_{P=0} \quad (2)$$

The *k*-values for *a*- and *c*-directions are (2.32±0.04)×10$^{-3}$ GPa$^{-1}$ and (2.12±0.07)×10$^{-2}$ GPa$^{-1}$, respectively. The $k_c/k_a$ ratio for *r*-BS is about 9 i.e. the largest among all studied A$^{III}$B$^{VI}$ compounds [24-29]. In other words, boron monosulfide has the highest anisotropy of the unit-cell compression in the family of A$^{III}$B$^{VI}$ layered compounds.

The pressure dependence of BS formula unit volume (V*) up to 50 GPa is presented in Fig. 3. Murnaghan (Eq. 3) [30], Birch-Murnaghan (Eq. 4) [31] and Vinet (Eq. 5) [32] equations of state have been used to fit the experimental data:

$$P(V) = \frac{B_0}{B_0'}\left[\left(\frac{V}{V_0}\right)^{B_0'} - 1\right] \tag{3}$$

$$P(V) = \frac{3B_0}{2}\left[\left(\frac{V_0}{V}\right)^{\frac{7}{3}} - \left(\frac{V_0}{V}\right)^{\frac{5}{3}}\right]\left\{1 + \frac{3}{4}(B_0' - 4)\left[\left(\frac{V_0}{V}\right)^{\frac{2}{3}} - 1\right]\right\} \tag{4}$$

$$P(V) = 3B_0\frac{(1-X)}{X^2}e^{\left(1.5(B_0'-1)(1-X)\right)} \tag{5}$$

where $X = \sqrt[3]{\frac{V}{V_0}}$, $V_0$ is unit cell volume at ambient pressure.

The values of bulk modulus $B_0$ and its first pressure derivative $B_0'$ for three considered EOSs are presented in Table II. It should be noted that all experimental data are in excellent agreement with our *ab initio* calculations (see Figs. 2 and 3). Rather high value of $B_0'$ can be explained by anisotropy of *r*-BS structure similar to those of graphite ($B_0' = 8.9$ [33]), hexagonal graphite-like boron nitride ($B_0' = 5.6$ [34]) and turbostratic BN ($B_0' = 11.4$ [35]). It should be noted that among all studied $A^{III}B^{VI}$ layered compounds [27,29,36-38] *r*-BS possesses the highest $B_0$-value i.e. boron monosulfide is the least compressible member of the $A^{III}B^{VI}$ family.

### B. *The new high-pressure phase of BS*

During compression the diffraction lines of *r*-BS monotonously moved towards the larger 2θ values. At about 35 GPa four new lines appear indicating the beginning of the phase transition towards a novel high-pressure phase of boron monosulfide, *hp*-BS (Fig. 4). Both BS phases coexist in the 35-50 GPa range, however, decompression of the sample results in the reverse phase transformation of *hp*-BS to *r*-BS.

According to the literature data, phase transitions in the 10-30 GPa pressure range are rather typical for $A^{III}B^{VI}$ layered compounds [1,3,36,39-41]. Nevertheless, the crystal structures of the new high-pressure $A^{III}B^{VI}$ phases have not been experimentally refined so far. According to *ab initio* calculations, NaCl-like structure is the most suitable and best matches with experimental data (e.g. for ε-GaSe [3]). However, all previously predicted dense $A^{III}B^{VI}$ structures (NaCl-like, CsCl-like, etc.) do not allow describing the experimental X-ray diffraction patterns of new high-pressure phase of boron monosulfide.

In order to solve the crystal structure of *hp*-BS we performed a set of *ab initio* calculations using the USPEX algorithm. According to our findings, the unit cell of *hp*-BS has trigonal symmetry and belongs to the space group *P*-3*m*1. It contains 1 independent boron atom (in *2c* Wyckoff position) and 1 independent sulfur atom (in *2d* Wyckoff position). The calculated difference in enthalpy between rhombohedral BS and new high-pressure phase ($H_{R-3m}$ - $H_{P-3m1}$) as a function of pressure is presented in Fig. 5. Below 34 GPa this difference is negative, and *r*-BS (*R*-3*m*) structure is more stable, while at higher pressures difference becomes positive which is indicative of a higher stability of the *P*-3*m*1 structure.

Rietveld refinement of powder diffraction pattern collected at 46.3 GPa was performed using Maud software [11]. The background was approximated by the 5-order polynomial. Structural information on *hp*-BS at this pressure is presented in Table III. As one can see from Fig. 6, the theoretically predicted crystal structure of *hp*-BS is in good agreement with experimental diffraction pattern except of three week lines that cannot be attributed to any known phase of the B–S system. Since these lines were observed before the phase transition (the first line appeared already at ~10 GPa), they cannot be attributed to *hp*-BS phase. We assume that the presence of these lines in combination with "wavy" background may be an indication of stacking faults that are quite expectable for the layered structures under quasi-hydrostatic compression. The final reliability factor $R_{wp}$ was converged down to 4.6%, which indicates a satisfactory quality of refinement, especially, for a boron compound.

The phase transition does not lead to any significant structure change; on the contrary, the high-pressure phase of boron monosulfide retains the layered structure built from the same trigonal $S_3B-BS_3$ antiprisms. As compared with the initial phase, the layers of the new structure are just shifted along the *b* axis, so the pairs of boron atoms in *hp*-BS are located on the parallel lines (opposite to each other) (Fig. 1*b*). In other words, the A-B-C stacking sequence in *r*-BS changes to the A-A-A stacking motif in *hp*-BS. According to our *ab initio* predictions such new layer arrangement makes *hp*-BS an electrical conductor, while the initial *r*-BS is known as a wide-gap semiconductor ($E_g$ = 3.4 eV [4]). The pressure dependencies of band gap energies for both BS polymorphs from electron density of states calculations using density functional theory with optimized exchange van der Waals functional are shown in Fig. 7. From this one can conclude that *r*-BS-to-*hp*-BS phase transformation at 34 GPa is accompanied by an insulator-metal transition.

Using Le Bail refinement routine implemented in PowderCell software [10], we determined the unit cell parameters of *hp*-BS in the 34-50 GPa pressure range (Fig. 2). The compressibility of *hp*-BS along the *c* axis was found to be slightly higher than that of *r*-BS i.e. the *c*-value of *hp*-BS is 3.4% less at 34 GPa and 5.9% less at 50 GPa than the corresponding *c**-values of *r*-BS, while the *a*-parameter of high-pressure phase is larger by 1.5% at 34 GPa and by 1.8% at 50 GPa. As can be seen from Fig. 3, (i) there is no noticeable volume change during phase transition, and (ii) *hp*-BS is slightly more compressible than *r*-BS that is somewhat unusual for a high-pressure phase. These both results are in good agreement with our DFT simulations (see Figs. 2, 3).

The observed phase transition in boron monosulfide, which is a "sliding" of one layer relative to two others, is very similar to the reversible first-order phase transition *β*-GaS → *ε*-GaS observed earlier at about 2 GPa [42]. According to X-ray diffraction data, the space group of GaS remained the same *(P6₃/mmc),* however, Ga atoms change their position from the 4*f* site (2/3,1/3,z) to the 4*e* site (0,0,z). This phase transition was also registered by discontinuous changes of Raman frequencies. Unfortunately, there is no literature data on compression of *ε*-GaS phase, which does not allow us to compare compressibilities of GaS polymorphs in order to check if high-pressure phase of GaS is also more compressible with regard to the low-pressure phase.

## V. Conclusions

300-K compression of rhombohedral boron monosulfide, *r*-BS, was studied *in situ* by angle-dispersive X-ray diffraction at pressures up to 50 GPa. The equation of state of *r*-BS was measured and the values of bulk modulus and its first pressure derivative were calculated by the fitting of experimental *P-V* data to Murnaghan, Birch-Murnaghan and Vinet EOSs. *r*-BS was found the least compressible member of the family of $A^{III}B^{VI}$ layered compounds. Formation of a new metastable high-pressure phase of boron monosulfide, *hp*-BS, was observed above 34 GPa. As it follows from electron density of states calculations, the phase transformation is accompanied by an insulator-metal transition. Based on *ab initio* evolutionary crystal structure predictions and Rietveld refinement of the experimental high-pressure X-ray diffraction data we found that *hp*-BS also has a layered structure, but with a different layer stacking sequence and belongs to the space group *P*-3*m*1.

## Acknowledgements

The authors thank Gilles Le Marchand and Dr. Alain Polian (IMPMC) for help in DAC preparation; Drs. Paolo Lotti and Boby Joseph (Xpress, Elettra) and Dr. Volodymyr Svitlyk (ID27, ESRF) for assistance in high-pressure experiments; Dr. Thierry Chauveau (LSPM) for help in Rietveld analysis, and Dr. Frédéric Datchi (IMPMC) for stimulating discussion. X-ray diffraction studies were carried out during beam time allocated to Proposal 20160061 at Elettra Sincrotrone Trieste and beam time kindly provided by the European Synchrotron Radiation Facility. This work was financially supported by the European Union's Horizon 2020 Research and Innovation Programme under the Flintstone2020 project (grant agreement No 689279), and by Russian Science Foundation (grant 16-13-10459).

Table I. Lattice parameters (hexagonal setting) and unit-cell volume of rhombohedral BS *versus* pressure at room temperature.

| p, GPa | a, Å | c, Å | V, Å³ | p, GPa | a, Å | c, Å | V, Å³ |
|---|---|---|---|---|---|---|---|
| 0    | 3.0586 | 20.3708 | 165.04 | 22.1 | 2.9360 | 17.5918 | 131.33 |
| 1.8  | 3.0476 | 19.8286 | 159.48 | 22.7 | 2.9360 | 17.6414 | 131.69 |
| 2.6  | 3.0334 | 19.5124 | 155.49 | 24.1 | 2.9285 | 17.5155 | 130.09 |
| 4.3  | 3.0312 | 19.3002 | 153.57 | 25.3 | 2.9240 | 17.4422 | 129.15 |
| 4.4  | 3.0237 | 19.1859 | 151.91 | 25.9 | 2.9249 | 17.4857 | 129.55 |
| 6.7  | 3.0099 | 18.8535 | 147.92 | 26.0 | 2.9224 | 17.4068 | 128.74 |
| 7.3  | 3.0155 | 18.7173 | 147.39 | 27.9 | 2.9096 | 17.2915 | 126.78 |
| 8.3  | 3.0015 | 18.6811 | 145.75 | 29.4 | 2.8986 | 17.2413 | 125.45 |
| 9.8  | 3.0015 | 18.4569 | 144.00 | 30.3 | 2.9058 | 17.2841 | 126.39 |
| 10.2 | 2.9912 | 18.4870 | 143.34 | 31.4 | 2.8935 | 17.2150 | 124.82 |
| 11.5 | 2.9860 | 18.3838 | 141.96 | 33.6 | 2.8844 | 17.1251 | 123.38 |
| 12.4 | 2.9800 | 18.2948 | 140.70 | 35.2 | 2.8851 | 17.1275 | 123.46 |
| 13.1 | 2.9772 | 18.2473 | 140.07 | 35.6 | 2.8775 | 17.0665 | 122.38 |
| 13.2 | 2.9837 | 18.1944 | 140.27 | 36.3 | 2.8818 | 17.0959 | 122.95 |
| 14.2 | 2.9709 | 18.1409 | 138.67 | 37.2 | 2.8787 | 17.0663 | 122.48 |
| 15.8 | 2.9708 | 18.0078 | 137.63 | 38.0 | 2.8731 | 17.0052 | 121.56 |
| 16.1 | 2.9650 | 17.9788 | 136.88 | 39.3 | 2.8736 | 17.0104 | 121.64 |
| 16.5 | 2.9622 | 17.9539 | 136.43 | 42.0 | 2.8575 | 16.9402 | 119.79 |
| 17.2 | 2.9565 | 17.9094 | 135.57 | 43.3 | 2.8592 | 16.9871 | 120.26 |
| 18.0 | 2.9552 | 17.8375 | 134.91 | 46.3 | 2.8437 | 17.0476 | 119.39 |
| 19.5 | 2.9539 | 17.7956 | 134.47 | 50.4 | 2.8380 | 16.9200 | 118.02 |
| 20.6 | 2.9432 | 17.6886 | 132.70 |      |        |         |        |

Table II. Equation-of-state parameters of rhombohedral BS. $\chi^2$ is an indication of the fit quality (lower for a better fit).

| EOS | $B_0$, GPa | $B_0'$ | $\chi^2$ |
|---|---|---|---|
| Murnaghan | 46.9±1.2 | 5.8±0.1 | 1.34 |
| Birch-Murnaghan | 41.7±1.8 | 7.9±0.4 | 1.26 |
| Vinet | 42.2±1.4 | 7.6±0.2 | 1.26 |

Table III. Structure parameters of the predicted high-pressure phase of BS at 46.3 GPa.

| Space group | Lattice parameter, Å | Atom | Atomic coordinates (fractional) | | |
|---|---|---|---|---|---|
| | | | $x$ | $y$ | $z$ |
| P-3m1 | $a$ = 2.8950(2) | B1 | 0.000 | 0.000 | -0.351(7) |
| | $c$ = 5.3218(12) | S1 | 0.333 | 0.667 | 0.212(2) |

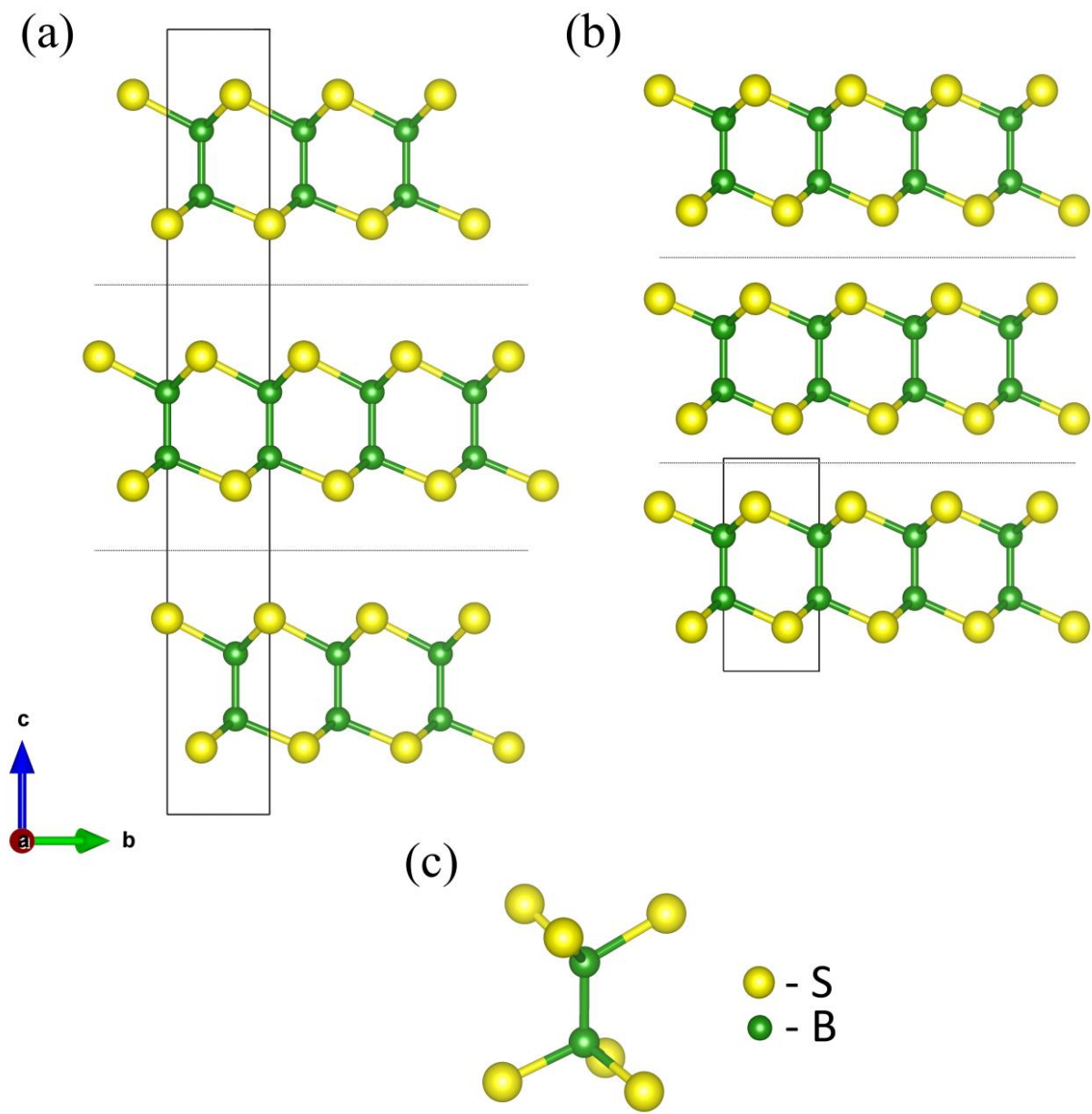

Fig. 1. Crystal structures of *r*-BS (*a*) and *hp*-BS (*b*) at 46.3 GPa; (*c*) common structural building block of both BS phases (boron atoms are green, sulfur atoms are yellow). The black rectangles show the unit cells.

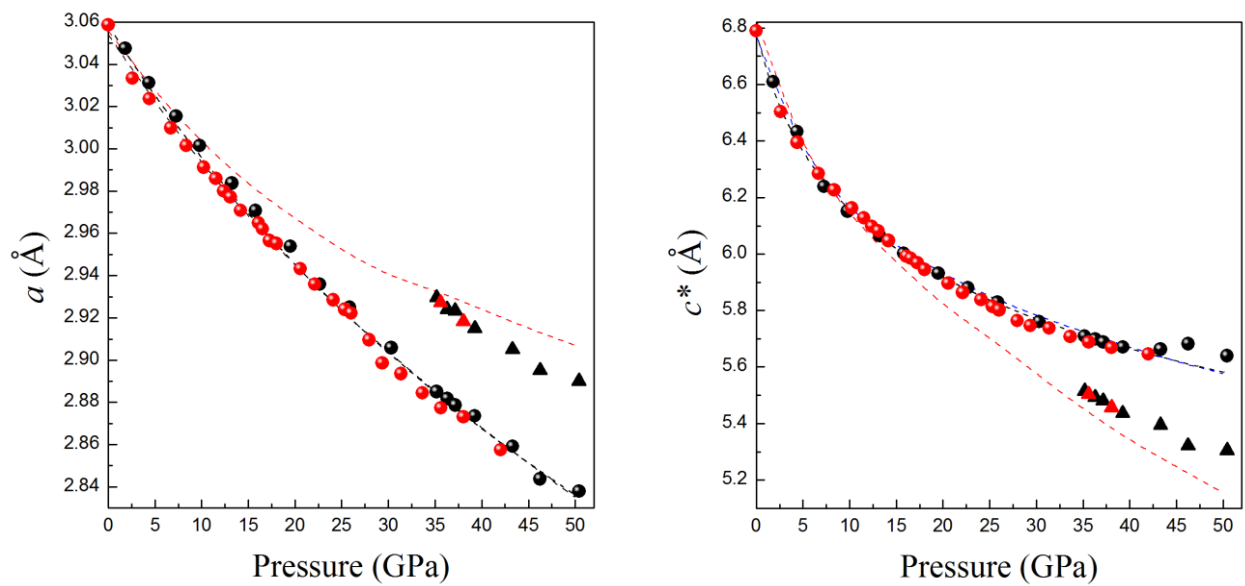

Fig. 2. The unit-cell parameters of *r*-BS ($c^* = c/3$ is lattice parameter normalized to one layer) and *hp*-BS *versus* pressure. The black and red symbols (circles for *r*-BS and triangles for *hp*-BS) represent experimental data obtained at Xpress and ID27 beamlines, respectively. The dashed black curves represent the fits of one-dimensional analog of Murnaghan EOS to the experimental data. The dashed blue and red curves show the corresponding USPEX theoretical predictions for *r*-BS and *hp*-BS, respectively.

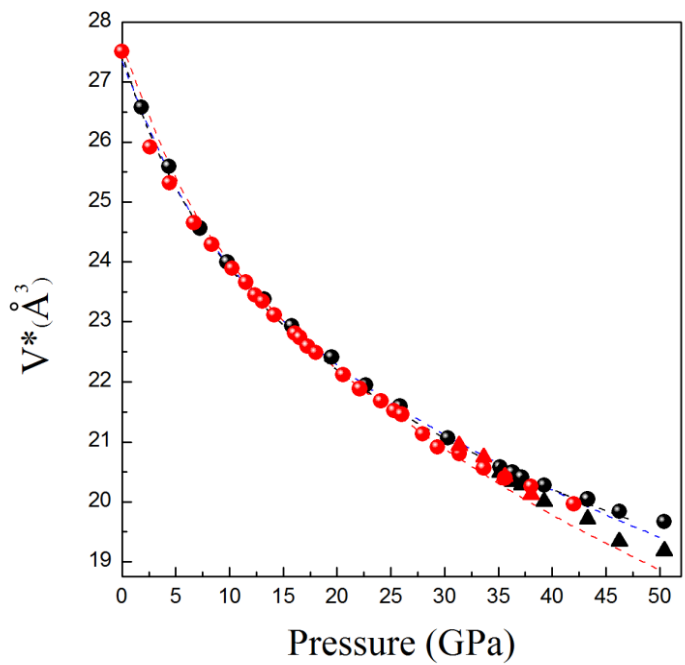

Fig. 3. The BS formula unit volumes (V*) of *r*-BS and *hp*-BS *versus* pressure. The black and red symbols (circles for *r*-BS and triangles for *hp*-BS) represent experimental data obtained at Xpress and ID27 beamlines, respectively. The dashed black curve represents the fit of Birch-Murnaghan EOS to the experimental data on *r*-BS compression. The dashed blue and red curves show the corresponding USPEX theoretical predictions of V*-values for *r*-BS and *hp*-BS, respectively.

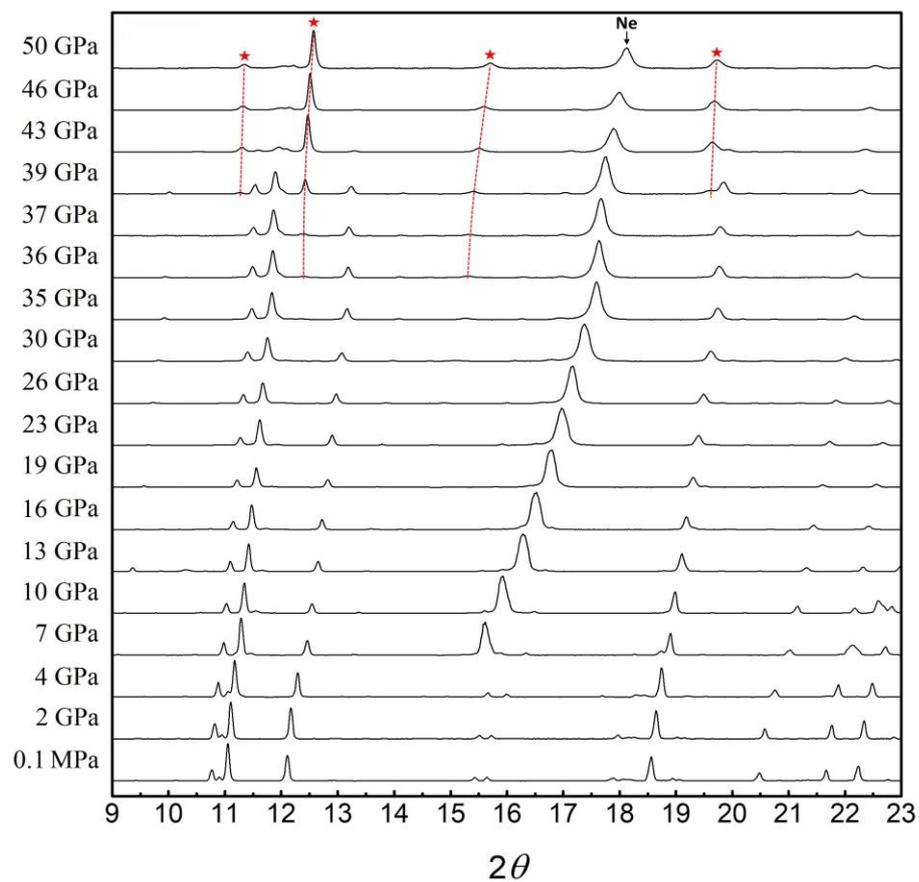

Fig. 4. 300-K sequence of X-ray diffraction patterns ($\lambda = 0.3738$ Å) of boron monosulfide collected at Xpress beamline (Elettra) in the 2-50 GPa range. The lines of *hp*-BS phase are marked by red stars; (*111*) reflection of solid neon is also indicated. Dashed red lines are guides for eye only.

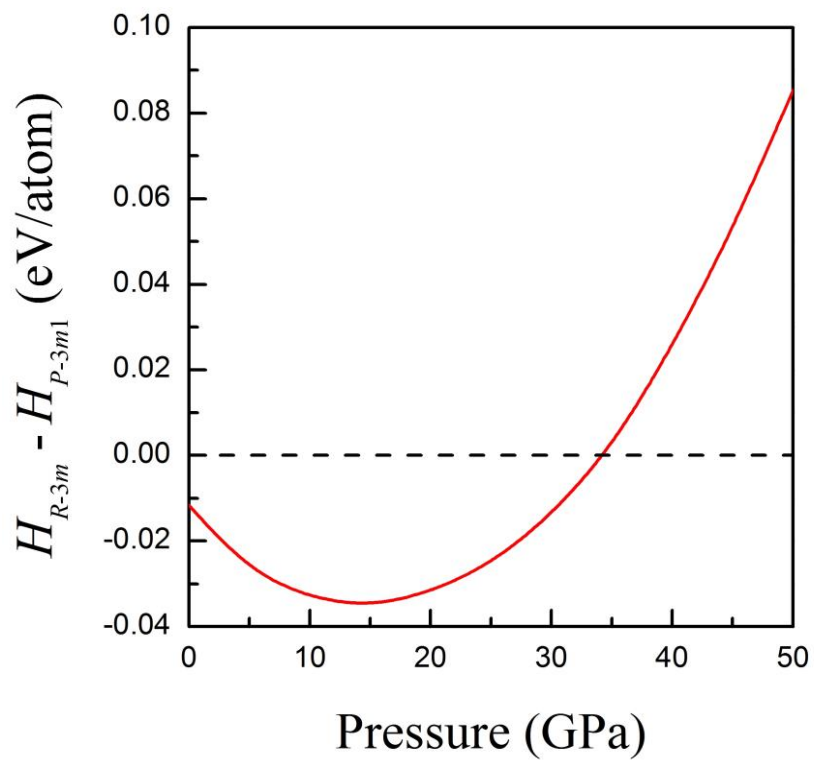

Fig. 5. Difference in enthalpy between *r*-BS (*R*-3*m*) and theoretically predicted *hp*-BS (*P*-3*m*1) phases.

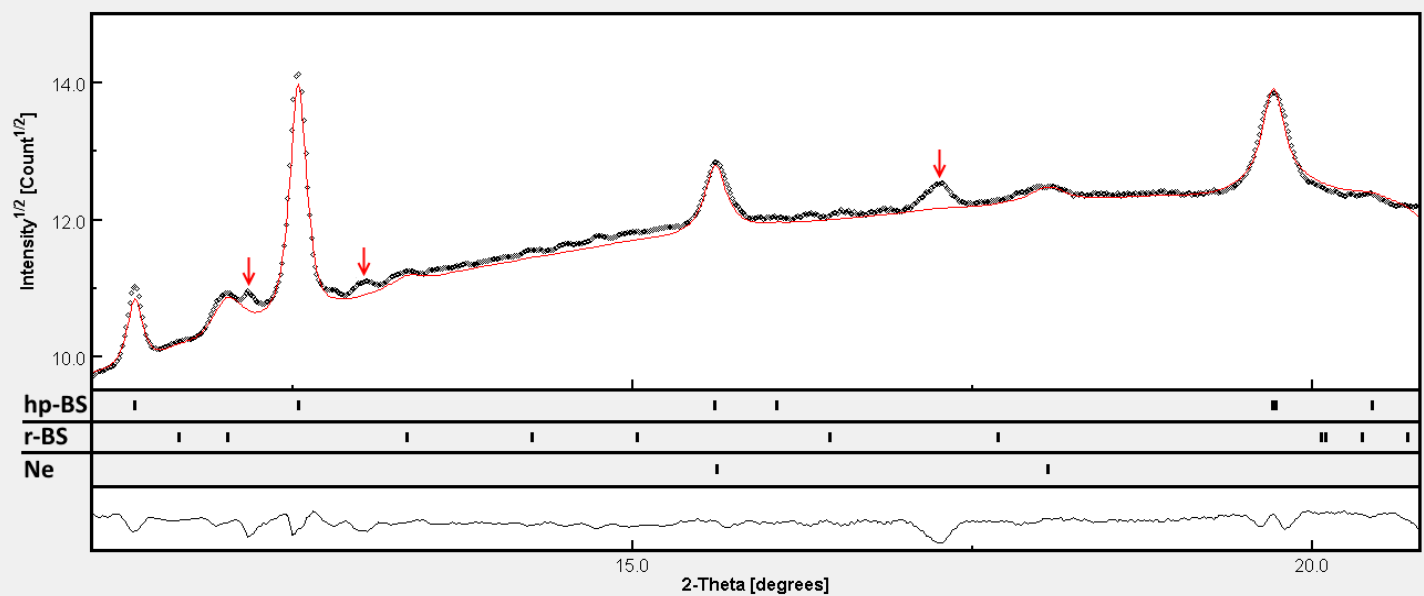

Fig. 6.  Rietveld full profile refinement of X-ray powder diffraction pattern of boron monosulfide at 46.3 GPa (non-attributed diffraction lines most probably caused by stacking faults are shown by red arrows).

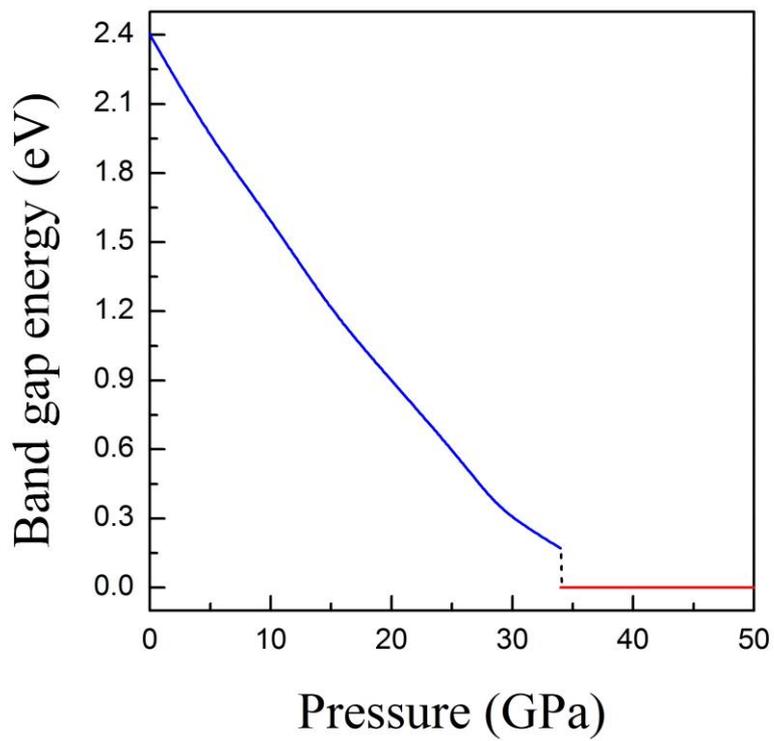

Fig. 7. Calculated pressure dependencies of band-gap energies for *r*-BS (blue) and *hp*-BS (red).